**Compositions in YBa2Cu3Oy for 3-D bond order superconductors**

H. Oesterreicher, Department of Chemistry, UCSD, La Jolla, CA 92093-0506

Self-doping in YBa2Cu3Oy starts to switch from a charge balancing of chain metal reduction and plane oxidation to one of chain ligand oxidation ($O^-$) and plane reduction on quenching from >600K. Amongst the responses are plane expansions and types of unusually strong superconductivity such as elevated temperature superconductivity (ETS), observed through laser pulsing (Tc=552K*) and upon shot quenching (Tc=200K*). We ascribe ETS to limited 3-D superconductivity due to a correlated system of bond ordering within chain-plane sandwiches and propose how to stabilize it. Accordingly, plane expanding n-doping arises from self-doping charge equilibration with local chain Cu of two-fold O coordination (2). The dumbbell type bonds of both apical $O^-$ are a result of high-energy environment and comparable in their metric with electron pairs on the plane. Following empirical Tc=%(2)x11 we suggest increasing %(2) to cause the observed retrograde rise in Tc (photoinduced reflectivity edge) to a theoretical limit of Tc=1100K* at 100%(2). We propose compositions and heat treatments based on charge-lattice commensurability. Paired charge concentrations on planes, determined by bond ordering based on magic number counts, suggest several promising candidates such as c=0.22=2/3a0x3b0, c=0.17=2/3x4 or c=0.080=2/5x5. The latter could be achieved with 32%(2) and display Tc=352K* not only in laser pulsed but in shot quenched materials.

## 1. Introduction:

A variety of properties of self-doped cuprate superconductors such as RBa2Cu3Oy=6+z (123 with R being rare earths or Y) find an explanation in their underlying nature as bond order materials [1,2]. This pertains to details in the doping curves such as linear portions or deviations thereof and especially critical charge concentration for onset, kink, plateau or optimal values. Charge-lattice commensurability of pairs in plaid arrangement dictates these critical compositions in magic number counts. As an example, the critical concentration for the optimum Tc in slow cooled YBa2Cu3Oy occurs around c=0.22 as seen in Knight shift or bond valence [1,3,4]. This corresponds to a plaid with 3a0x3b0 per pair, or c=2/3x3=0.222, obtainable at ideal z=4c=0.889 on bond order principles. Tendencies to a Tc plateau exist near c=0.167=2/3x4, corresponding to z=0.667. The critical kink location, above which all doped charge has been transformed into pairs, is c=0.080=2/5x5 observable at z=0.32. These slow cooled materials crystallize, from about z>0.3 on, with an orthorhombic structure based on local environment (n) with n=2 and 4 referred to as O24 (or orthoI). At lower z they can be obtained as semiconducting tetragonal T23. We take their cell volume type as V0, for purposes of calibrating for relative cell volume effects in other materials. O24, by comparison, shows relative plane and cell volume contractions (P-V-) while high temperature O3 modifications can display contrasting V+P+ effects with changes of doping from plane p to n-doping implied. The region near oxygen half-filling (123/6.5) is sensitive to transitions between these types.

These findings and ideas have gained renewed relevance in the light of discoveries of signs for elevated Tc superconductivity (ETS), observed upon quenching or partial re-oxygenation [5] (Tc=200K*) and laser pulsing [6,7] (Tc=552K*). With ETS, pair formation has been postulated to occur in addition to planes also on the 2-fold O coordinated chain site (2) [8]. An unusually low Cu-O-Cu distance between planes and chains supports this as it becomes comparable to the one within planes. Mixed plane electron and chain hole self-doping creates unusually high pair concentrations and limited 3-D superconductivity within the plane-chain-plane sandwich.

A relation of ETS also exists with high temperature modifications. Materials shot quenched (SQ) from >600K, or specially re-oxygenated, obtain with increasingly expanded planes and cell volumes (P+V+), optimizing around 670K. Optimized V+ are taken as n-doped on planes but without pairs on chains. V+ materials of O32 type show intrinsically slow kinetic (high activation energy) of transformation to the room temperature O24 modification. This has been explained as due to electronic rearrangement involving $O^-$ (subperoxides) [9], characteristic of V+. The concept of subperoxides, as charge sharing $O_q^{1-2q}$ aggregates, also explains inertness to O2 evolution in acid, structural trends, enhanced Tc potential and recalcitrance to oxygenate beyond y=6.7 in the manner of conventional materials. Spectroscopy has further corroborated this $O^-$ view [10] in high temperature modifications. The presence of subperoxidic $O^-$ is taken to represent a hallmark of a wide range of high temperature cuprates, in keeping with the related metal-peroxide stability belt under those conditions. Subperoxides may represent a salient feature for ETS as well. In the following we will elaborate on the literature germane to these effects as a basis for further materials design.

### 1.1. Bond order formalisms

The essence of the bond order concept is an arrangement of paired charge in an electronic crystal [1,2]. It was postulated in the form of a plaid structure that would have special stability at charge lattice commensurability [11-16]. Signs for an electronic crystal were indeed found in ARPES and TEM [17,18]. Given perpendicular plaid lines of doped charge with indications for pair formation at 3a0/2 distance on planes (in trijugate bond position or 3J), one expects special self-doping charge concentrations to show the hallmarks of the expected extra stability due to charge lattice commensurability. They run a series of magic number counts in multiples of lattice parameter such as c=2/Pab, where Pab corresponds to charge periodicity na0xmb0. Examples are the often observed optimal doping at c=0.125, c=0.17 or 0.22, which find a natural explanation as 2/4x4, 2/3x4 and 2/3x3 respectively. The self-doping in 123 is considered to reflect a competition for charge between planes and chain sites, adjusting through it their changing relative tension with temperature. The plaid structure dictates that ¼ of the potential dopant on chains will dope the planes. This leads to a linear relationship of self-dopant concentration to Tc in the range where all doped charge is in the form of pairs (source region or potentially linear portion of the doping curve of 123 from z=0.080 to 0.89). Where dopant (4) is linearly increasing with z in the source region so will be Tc.

ETS is here defined as Tc>150K. A limit near 150K is characteristic for predominantly 2-D cuprate superconductivity according to a linear Tc(c) (equation1). This relation is specified in the plane isolation model [2] based on a direct dependence of superfluid density on Tc according to

equation1a Tc=ic,

where a plane isolation factor, i, takes into account the deleterious approach of apical O to the planes. For infinite d, optimal i=600K. Heuristically, c=0.25 appears to be an optimal self-doped charge concentration [referenced in 3 and 4] in some plane p-doped cuprates with multiple charge balance layers, while it is confined to c=0.22 in 123 for reasons outlined further below. The model is general and extendible to other materials such as pnictides [2]. For the self-doped 123 indications are that self-doped charge is transformed into paired charge over a large range before over-doping. This includes part of the region of insipient orthorhombic splitting. It manifests itself in a relationship with changes in lattice parameter c or parameter d, representing the distance of plane to apical O. For 123, the relatively small size of the deleterious d diminishes the isolation factor to about i=400K. Bond valence or an empirical $i=73d^2$ with d in A allow for calculations. A dependence of Tc on d appeared also on hybridization calculations [19] but no predictive scheme was developed as a result.

As (4) is held solely responsible for doping balance with planes in conventional materials, it happens that the linear Tc(c) relationship can also be expressed as

equation1b Tc=%Dxconst

where percent dopant %D= %(4), with empirically determined const=1.0 and direct relations with equation1a. Both forms of the linear Tc(c) relationship hold from z=0.32 to 0.89. This corresponds to Tc=32K to 89K (careful preparation can slightly increase these values).

Equation1b can include other dopants, with const being 1.1 for (3) and 11.04 for ETS based on (2). Linear dependence of Tc on superfluid density appears as the hallmark of self-doped systems extending to a yet undetermined lower concentration. The conventional parabolic model as a theoretical mainstay is not representing the doping curve shape and is misleading for self-doped systems [20].

### 1.2. High temperature materials

Several high temperature techniques have routinely produced strong increases in structural parameters such as cell volume or plane metrics around O half filling of $RBa_2Cu_3O_y$ (123/6.5) [21-27]. They include direct SQ from temperatures >600K to 950K or partial re-oxygenation of thermally sequenced materials around 670K. As an example, the V+P+ effects for $YBa_2Cu_3O_{y=6+z}$ can be observed on SQ with small sample size of 0.2g, shortening quenching time to 0.1s. In addition to increasing quenching speed, increases in activation energy of the transition reaction to low temperature modifications can be engineered through chemical means (partial substitutions). Under those circumstances indications for strong cell volume and plane expansion effects (V+P+) appeared for several rare earth and Y analogs near experimental y=6.46, which was assumed to reflect an ideal plane bond order at y=6.444 with doped charge concentration on planes of c=0.222. The V+ effects maximized at 670K with volume increases of up to 2.7% over V- [5]. In some instances they were coupled with a doubling and quadrupling of Tc.

An unidentified minority component of the above preparations with Tc=200K* level is here related to ETS. In addition, around 1050K a phase with strong plane expansion and c-axis contraction but volume corresponding to the semiconductor (V0) was found and designated C-phase of C-V0 type [28,29]. It is also discussed as related to ETS. Findings concerning unusual high temperature phases were an outcome of investigations of the effects of partial transition metal substitutions for Cu. The resulting micro-cluster formation [30-39], leading to some reductions in Tc and rise in magnetic remanence, was also found to reduce the speed of transformation reactions from the metastable high temperature O3 modifications to the conventional O42 orthorhombics. Also on partial Al substitution for Cu, chain length was reduced [40].

Various O-ordering on chain sites has been discussed in terms of O-O potentials for patterns such as Stripes (O24), Argyle (O3), Herringbone (T23) or further varieties [41-47]. The general thermodynamic metastability of 123 versus temperature involves the more stable decomposition into peroxides through their higher overall O-uptake [48,49] in the peroxide window of stability. This higher overall O-uptake is the crux in developing P+ phenomenology based on subperoxides. T23 is a tetragonal structure with disordered placement of non-doping (3) that can on slow cooling develop into self-doping O24. It can however also transition into the orthorhombic Argyle (O3) structure. In this pattern, subperoxides can form at an increased O content compared to the conventional. These materials can become electron self-doped on planes. General phenomenology of electron-doped superconductors has been reviewed in [50] and for cuprates in general in [51].

### 1.3. Laser pulsed materials

One can see laser pulsed materials as creating structural properties of a thick family of compounds. This irradiation drives [6,7] a conventionally prepared $YBa_2Cu_3O_{y=6.5}$ ceramic with Tc=52K to one showing coherent interlayer transport strongly reminiscent of superconductivity up to Tc=552K*. Accompanying the electronic transition is a structural one: the 5-fold O coordinated copper dioxide double layers, denoted here as (5) become more separated by two picometers as well as more buckled, and the layer distance between them becomes thinner by a similar amount. Clearly, laser pulsing has resulted in transforming the structure type from a thin family with respect to axial ratios with c/p>1 (p=3(a+b)/2) to hallmarks of the thick family denoted as C-P+ with c/p near 1. Another member of the thick family is a tetragonal or pseudo-cubic phase with a=c/3=3.87A, denoted the C- phase [28,29]. It may hold a key to understanding ETS.

We assume that laser electronic excitation leads to a switch of doping type on (5) to n-doped due to the plane expansion as a transient structural linkage between thin and thick family hallmarks. In addition, pair formation of p-type occurs on the chains leading to 3-D coupling effects by an order of magnitude stronger than known so far. It has been proposed that this involves local environment (2). Calibrating to the value of Tc=552K* one can, for orienting purposes within ETS, empirically expand equation1a to 3-D effects with equation1b Tc=%(2)11.04 over some bond order dictated region. In this formulation only contributions from (2) are accounted for. The question concerning charge concentration leaves some ambiguity as dealt with further below.

## 2. Results

We will in the following make a distinction between a selection of materials for laser pulsing and for SQ of $YBa_2Cu_3O_y$ respectively. Laser pulsing of selected materials with different y and mix of (n) is mainly of academic interest for learning about the sensitivities of ETS to these parameters, albeit with wide boundaries for exploring possibilities. By comparison, SQ can be used to obtain practical materials for further studies but it stays limited with severe restrictions in synthesis. Both methods have in common a quest for establishing elements of a thick family of compounds with C-P+ effects, which are held promising for attaining ETS. This translates into unraveling the counter play of maximizing and stabilizing C-P+ effects through the balancing of influences or outright dominance of either (3) or (4) with the different resulting % of (2) that appears to be important for the strength of ETS. Structural and electronic types in $YBa_2Cu_3O_y$ and comparison compounds, indicating 3 categories P-, P+, and ETS, are collected for perusal in table1. Selected predictions concerning particular materials are collected in table2. The aim is here and in the following of showing how at a given y, different mixes of (n) can obtain with different potential for the strength of ETS. When a C-P+ material is stabilized with ETS properties, we assume that its strength is related to %(2), setting a theoretical limit at Tc=1100K* for 100%(2).

### 2.1. Laser pulsing opportunities

We will elaborate on factors involved in ETS phenomenology to apply it first to laser pulsing. For this purpose we deal with the question concerning the corresponding charge concentration due to %(2) in equation1 that has left some ambiguity. In this respect we need to know how the assumed extra pairs on chains add to the pairs on planes. If for laser pulsing of 123/6.5 a comparable doping count exists on planes as with the conventional material with c=0.125, then we have 0.25 paired charges on chains and a total of c'=0.50 per formula unit (fu). We refer to this as the doubling counting scheme. It simply states that an equivalent number of pairs will now reside also on chains through doping balance with planes. In a straightforward way this might imply a doubling of Tc, rather than the observed 11-fold increase. One could then account for the discrepancy as a result of factors connected with the different local pair concentrations and 3-D effects. Those could include the doubled pair concentration as counted per Cu on chains and its interaction with the planes.

Generally, we consider that an electronic crystal exists not only on planes but also in a corresponding manner on chains. Accordingly, we assume that similar channels of self-doped charge exist on chains separating within a plaid pattern. Translating Tc=%(4)x1.0 for conventional materials into this picture one notes that within the doped charge channels every second (4) has to be doped by stoichiometry. The same alternate metal doping scheme holds when counting for (3) with V+ materials although the actual charge carriers appear here to be subperoxidic O within the doped charge channels. From these considerations one also understands limits of optimal doping that differ between one and two charge balance layers (c=0.22 vs. 0.25) due to doping exhaustion.

Within this view it is now possible that alternative phenomena come into play for ETS. One such possibility is that sticks of n-doping (2) may actually have been keeping a similar self-doping geometry on the plaid grid as with the original p-doping (4). However, now there are two doped charges per (2) due to the 2 apical $O^-$ dumbbells. A natural expectation would then be a quadrupling of paired charge concentration counted per formula unit. We refer to this as the quadrupling counting scheme. The total paired charge concentration counted per plane Cu is then c=0.25 and c'=1.0/fu. On using equation1a in the form Tc=600c'=600K* one sees that the 2-D formula can provide rough estimates for the magnitude of Tc in this counting scheme for ETS, assuming i=600K.

However, for practicality we proceed by using equation1b Tc=50%(2)x11.04=552K* as an empirical calibration and consider the electronic crystals, corresponding to %(2) or c, on the doubling counting scheme. Accordingly we estimate a theoretical limit for ETS as Tc=100%(2)x11=1100K*. This could be obtainable through laser pulsing the semiconductor with z=0.0. However, while thought provoking, this limit may not be achievable. Practical limits will likely be set by bond order considerations and a mutual support of the various (n) in supporting the existence of a thick compound (C-P+) with small distance of chain to plane, d1+d.

We will now search for such stabilizing criteria. Close to breakdown of orthorhombicity, one can predict, with Tc=%(2)x11, a conventional material with z=0.333, 66.7%(2) and 33.3%(4) to show Tc=734K* on laser pulsing. By contrast, the conventionally prepared material has Tc=%(4)x1.0=33K in line with experiment. On the doubling counting scheme c=0.167=2/3x4 on the plane, in line with extra stability through bond ordering at these charge concentrations for ETS. The remaining 33%(4) are assumed to be doping-inactive as with the analog z=0.50 on laser pulsing and potentially stabilizing the thick structure type. The assumption of pairs on (2) therefore produces a trend, counter to conventional, as Tc decreases with z and optimal Tc are expected at bond orders corresponding to low z. A retrograde tendency has also been reported for the photoinduced reflectivity edge on laser pulsing [7], giving support to our assumptions.

If materials in the semiconductor range can be similarly excited with lasers one could expect z=0.056 to produce the c=0.222=2/3x3 bond order. This corresponds to 88.9%(2) and a calculated Tc=981K*. On the lower %(2) side the conventionally prepared material with z=0.68 has 32%(2), and accordingly one calculates Tc=352K*. The high concentration of remaining 68%(4) may preclude its being doping-inactive, though. 32%(2) was chosen because on doubling counting scheme it represents a bond order according to c=0.080=2/5x5 on planes. Interpreting this material on quadrupling counting scheme one finds bond order at c=0.16=4/5x5. Also c=0.167=2/3x4 on planes is close by. Analysis of similar cases will allow a distinction between the 2 counting schemes.

### 2.2. Shot quenching opportunities

Quenched materials can be of interest per se or as materials that are subjected subsequently to laser pulsing. SQ demonstrated V+P+ effects for O32 with ideal z=0.444 preparing the stage for the assumed plane n-doping of ETS. Laser pulsing of this material could transform the 11%(2) into ETS with Tc=121K*. Also, one notices an extra dimension to the richness in phenomenology through the capability of O32 to gradually transition to conventional V- O24 as a function of time and temperature. While the ideal parent material has 11%(2) and 89%(3) it will gradually transform and eventually reach 56%(2) and 44%(4). The latter corresponds to calculated Tc= 616K* on laser pulsing. Within this transition the material apparently can run through one or several bond orders on planes and so potentially reach conditions for extra stability for ETS to be observed in the bulk, before it will arrive at conventional V-P-. Intermediary examples are materials with 22%(2) with calculated Tc=242K*. On the doubling counting scheme this corresponds to charge concentration on planes of c=22%(2)/400=0.055=2/6x6. This could represent the experimental indication for a Tc=200K* level on SQ. The transition should also reach 50%(2) with a Tc=550K* level, corresponding to c=0.125=2/4x4. It may be difficult to stabilize P+ effects at relatively high %(2) though. Similar transitions can be expected at other materials with P+V+ effects.

On the rules outlined, several other possibilities for ETS exist on SQ. We discuss here the z=0.32 material with 68%(3) and 32%(2). It forms an interesting counterpart to the conventionally prepared z=0.32, as outlined above with its 68%(2) and 32%(4). If only the 32%(2) are induced to become doping upon laser pulsing, then Tc=352K*. This would, in the double counting scheme, correspond to 0.080=2/5x5 on planes. As d of doping inactive (3) is perhaps not an obvious choice in supporting the oxidation of (2) it remains an open question whether it could be successfully employed as a stabilizer of ETS.

The same reservation concerning the usefulness of (3) in stabilizing C-P+ pertains to other materials in the semiconductor range. As an example, z=0.25 should be comparable with respect to harboring 50% (2) to the conventional z=0.5, but now with (3) making up the rest and potentially participating in the doping and leading to the 550K* level. Other bond orders are of interest such as z=0.16 with 67%(2) and c=0.17=2/3x4. It is even possible that the rather stable and non-superconducting C-P+ phase at z=0.67 could become superconducting through laser pulsing due to unfreezing of the presumed localization catastrophe. As it probably has no (2) it can only reflect the potential self-doping of (3) and as such may have comparable Tc=67K with conventional material.

Besides several high temperature modifications, another opportunity for material engineering is the low temperature equilibrated region of tetragonal to orthorhombic transition. With increasing z one has here, in the insipient orthorhombic, another possibility to tailor the mix of (n). However, in this range P- effects set in early in the development of (4) and external parameters such as pressure would have to turn the doping around to P+. Laser pulsing can of course shine more light on the phenomenology of this potentially rich responsive range.

**3. Discussion**

The encompassing view of several high temperature modifications in 123 near y=6.5 as based on self-doped $O^-$, presumably in a subperoxide state in conjunction with (3), explains their unusual behavior. This includes the relatively slow transformation of O32 to low temperature modifications O24, as due to the need for severe electronic rearrangement on chains from ligand based self-doping to metal based. This view also explains pronounced P+V+ effects and strength of Tc, apparently connected with changes in doping type to plane electron doping near O half filling. In this state they can exhibit a Tc=100K level. These subperoxidic materials are also inert to oxygen evolution in acid. An ideal composition for a representative phase is taken as y=6.44 on bond order principles. It is obtainable by either quenching and/or re-oxygenation in the window from 670K to 950K. In addition subperoxidic materials are recalcitrant to full oxygenation beyond y=6.7, reverting to non-superconducting at their limit of oxygenation. Around this limit they can be of O34 type or of the T34 type of the C- phase. This combination of properties sets them apart as a yet not sufficiently recognized doping variety in cuprates.

Subperoxides remind of thermodynamic metastability of 123 versus temperature in the metal-peroxide window of stability where the latter offer higher relative O-uptake compared to conventional preparation without decomposition. Subperoxides can form within the O3 pattern of preferred orientation of O placement. They constitute the self-doped region on chains that is reflecting the geometry of the electronic crystal known for the planes. Their genesis is connected with a critical dependence of cluster size of (3). Once this Argyle (O3) type structure with its subperoxide channels is formed under electronic rearrangement out of semiconducting T23, with patterns such as Herringbone, a gating is set for it to stay in type. This leads to resistance to a change to the electronic rearrangement needed for Stripes of (O24) with its corresponding structure of self-doped regions based on reduced (4). However, while activation energies for this transformation are relatively high, the elevated temperatures in question can still result in challengingly fast transformation rates from a synthesis point.

In a similar vein one can expect ETS materials to also be based on $O^-$. Laser pulsing indicates a P+C- type and that is also assumed for a Tc=200K* level in an unidentified SQ minority. In the ETS case, however, we assume that two apical $O^-$ be connected with (2), creating a dumbbell configuration with pairs on chains. This multiplies pair concentration and introduces predicted 3-D effects as a result of increased connectivity perpendicular to the planes [52]. At high temperatures planes can become even more oxidizing of chains (robbing them of electrons). This phenomenon at its extreme can happen with (2) as it can formally undergo oxidation by two steps from 1+ to 3+, although self-doped charge will reside on oxygen. Further criteria for stabilizing these modifications are seen as connected with charge-lattice commensurability concerning self-doped charge in magic number counts on the planes with potential correlations with chains. In addition, advantageous competition of local environments within chain sites for the distance to the planes will predispose to the severe C- effects. We have dealt with the C- phase as a close, non-superconducting structural relative, possibly based on peroxidic O. It is stable around 1050K and we expect that this is also the range for ETS material stability if perhaps somewhat lower. Accordingly synthesis at these elevated temperatures again poses challenges, asking for special techniques concerning extremely fast quenching and transformation retardation through partial substitutions (e.g. Ni for Cu).

By comparison, the conventional low temperature equilibrated materials are based on over-oxidized metal in (4) coordination rather than similar effects involving its ligand. This metal, on chain sites, is self-doping the planes, now without $O^-$. In the process of hole doping, planes display attendant P- effects, and chains are reduced partially on their high positive metal charge centers. These materials readily evolve oxygen in acid, a process that is useful in titration of (4). The range of onset of orthorhombic splitting with y, with its rise of (4) and decline of (3), represents a versatile range and experimental arena in its own right to explore complex behavior. In fact, as it can also spawn ETS under laser pulsing, (4) promises to be capable of stabilizing bulk ETS materials based on (2). It will be interesting to witness whether the subperoxidic type at high temperature with (3) phenomenology or else its low temperature (4) counterpart will win the supposedly achievable prize of leading first to room temperature Tc as a relatively stable 3-D phenomenon in the bulk.

### 3.1. Apical oxygen and Tc

We will now further discuss our predictions concerning optimal compositions of high temperature modifications on bond order dictated constraints. As outlined, the question of pair number or charge concentration in assumed 3-D ETS can be approached through the generalized formalism of equation1 Tc=ic =%Di/400. A straightforward comparison between p-doped and n/p doped with i/400=1.0 and 11 indicates a more than 10 fold increase for the latter. Arguments were given that 3-D effects are likely to be involved in this increase.

In principle, systems with apical O are not known for tending to be electron doped on planes. Rather, decreasing distance of the apical to the plane is generally seen as deleterious to Tc. However, in high-energy environments this restriction appears to be relaxed and planes will expand. This can happen in a first step to P+ states involving (3). In a second step the novel situation of 3-D superconductivity can come into play based on (2). For ETS one assumes now that the apical O will display strongly decreased values of d+d1, comparable to the one of the Cu-O distances in the planes. This is therefore not a conventionally electron doped system with apical O, as we assume the 100K level of V+/O32 to be, but rather represents a novel situation of resonating 3J bonds. The gist is that YBa2Cu3O6+z can switch to either only n-doped, or else ETS with n-doped planes and a chain 3-D component. In the latter the apical O is not a hindrance but an active participant in 3-D superconductivity. In fact, it will be of interest to investigate the influence of d of parent compounds in predisposing them to become ETS materials. The susceptibility of planes to either doping type apparently also depends, besides on the supporting chain structure, on the options concerning bond ordering on the planes.

### 3.2. Bond-ordering phenomena

Generally, one understands bond-ordering phenomena as related to structural adjustment measures and so to deeply connect with crystallography. Bond-ordering phenomena support changing layer fit requirements as a function of temperature. As such this can lead to changes in doping type as a function of thermal history and therefore to qualitatively different doping curves. The 2 extremes of doping curves depend on the location of equilibrium1 (2)+(4)= 2(3). Equilibrating at low temperatures will shift equilibrium1 to the left, creating the conventional behavior based on (4) with Tc optimum at theoretical z=0.889 and c=0.222 and properties of a thin structural family (high c/p). Equilibrating at high temperatures will shift equilibrium1 to the right and create behavior based on (3) with Tc optimum at theoretical z=0.444 and c=0.222. The respective doping curve can then be assumed to represent the essential features of the one corresponding to V- types by condensing it within half the range in z and inverting the doping type. A presumably partly ionic over-doped non-superconducting compound (C- phase at z=0.67) follows this as a representative of a thick family of compounds. We assume that potential ETS materials will also show aspects of this type, albeit at lower z.

Between those 2 extreme doping curves one can assume rather complex mixed behavior as a function of changing heat treatment. Also, transition from (3)-phenomenology to one based on (4), on equilibrating a particular preparation near room temperature or in a range without further O-uptake, allows an expanded opportunity to develop special behavior at a given z. Several interesting comparisons follow from those 2 extreme doping curves. The analog of the laser pulsing z=0.5, based on (2) and (4), would be z=0.25 with 50%(2) and 50%(3). This analog has a chance to exhibit on laser pulsing a similar Tc=550K* level due to identical 50% (2). In this case, however, the presence of (3) may contribute doped charge and help to stabilize the P+ effects so that ETS might already be observable in a quenched material. It might help if one could extend orthorhombicity of O3 type to this composition.

The beginning of the linear portion in Tc with c (the source-region) is near z=0.32 for conventional materials, corresponding to c=0.080=2/5x5 due to 32% (4). For (3)-phenomenology one predicts this to happen with half the above at z= 0.16 with identical c=0.080 due to 32% (3). The complementing 68% (2) has therefore a potential for ETS with Tc=748K* either on quenching or laser pulsing.

A drastic difference pertains to z=0.67. On conventional (4)-phenomenology this composition is the parent of the Tc=67K plateau region with its c=0.167=2/3x4 at 67% (4) and 33% (2). Laser pulsing may show Tc=363K*, although the large 67% (4) may preclude this. On quench or (3)-phenomenology, z=0.67 corresponds to the non-superconducting C-P+ phase with 67% (3) and 33% (4) and it is not clear whether laser pulsing can have an un-freezing effect on the underlying localization catastrophe. This C-P+ phase is perhaps the most enigmatic piece in the puzzle of ETS as it has the structural signature of it but Tc<4K. In this connection it will also be interesting to detail the difference between the latter non-superconductor and the one observed in the transition of O3 to O42. For the O3 to O42 transition a typical material would have 26%(3), 42%(2) and 32%(4). A competition between (3) and (4) appears to be deleterious to superconductivity. However, (4) can act as a plane dopant beyond z=0.222 or c=0.0555=2/6x6 in the presence of (2). By comparison, local environments (3) can behave either as a dopant or non-dopant in the presence of (2), the later being characteristic of the semiconductor range. So far (3) has been shown to become a dopant near 89%, leading to V+ at z=0.44. However, V+ effect have been observed also at selected materials with z<0.44, in both orthorhombic and tetragonal symmetry, so that a critical dopant concentration is not yet established.

### 3.3. Competition between (n)

For obtaining ETS and particularly Tc>300K* in quenched materials, one notes that compromises will have to be reached between 2 influences. There is the need for large % (2) to gain high ETS but in this case one is apt to loose P+ effects through the presumably needed support by (3) or (4). If one is satisfied with lower % (2) one doesn't only decrease ETS but chances are that other components will now dominate in their doping behavior and preclude (2) from attaining ETS or lead to a localization catastrophe altogether.

It is tempting to introduce further parameters, beyond temperature or its equivalents in laser energy, into the play of self-doping charge balancing. One could be the application of external pressure along the c-axis. It is not unlikely that pressure could crush the apical O closer to the planes even in conventional materials. This could result in the yet elusive 3-D superconductivity counterpart to plane p-doped conventional materials. The V+ O32 could be induced to become C-P+ although the small %(2) is not expected to increase Tc strongly. External pressure could also induce switches between P+ and P- types when applied in the plane direction. An intriguing challenge would be to pressurize a single crystal of material with y=6.0 along the c-direction. This could result in the transition of a semiconductor to a 3-D superconductor with Tc=1100K* due to the forced interaction of (2) sticks with a plane capable of expanding. Putting pressure parallel to planes could induce (3) to become reducing to planes and so show an ambivalent nature as both an oxidizing and reducing dopant.

### 3.4. Outlook

The actual spatial arrangements of charge and their superconducting transport are open to conjecture. In one picture, the charged chain sticks are penetrating to the electron-doped planes and be participating in bonding. In fact, if a direct 3-D propagation of these expansive n-doped and contractive p-doped pairs occurs by dynamic exchange of them, then an averaging of their effects would be expected, so that from a phase stability point these materials would resemble aspects of the semiconductor they are derived from. This return to the stability criteria of the parent material may in fact be reflected in the considerably smaller cell volume expansion effects, compared to shot quenching, estimated at 0.9% on laser pulsing from the limited data in [6,7]. This assumed dynamic pair exchange may be obviating the need to further distinguish the doping type on specific locations as pairs may be exchanged. It is also possible that 3-D charge density wave properties come into play as recently determined [53] for z=0.67.

Alternative mechanisms of charge transport come to mind. One would involve mobile peroxidic units formed from the apical O of two adjacent sticks of (2). It is then possible that this assumed charge transport is directly coupled with the one on planes.

The magnitude of Tc=552K* of laser pulsed materials and the potential for reaching Tc=1000K* invites musings on preparative complexity. O diffusion and rearrangement generally start to take place in a noticeable fashion in the range of 550K. A straightforward attempt at synthesizing this Tc=552K* material by increasing temperature, could accordingly fall short because of the shift in equilibrium1 to more (3), destroying the conventional material that supports ETS. The state corresponding to Tc=552K* appears to be capable of maintaining only a fleeting existence before it would rearrange to incorporate (3) and it is not clear how much of it would be tolerated for ETS. The range 900-1100K, on the other hand, is noteworthy as a temperature window where a main part of the P+ phenomenology is set in track. Here the situation may look more hopeful for stabilizing ETS, as it is possible that some materials, heated to this temperature, may actually transform to become superconducting near or at the synthesis condition if perhaps only in a transitory manner.

Self-doping leading to electronic crystals with plaid type patterns may be at the heart of phase stability questions in a more general way. It may be ubiquitously leading to the accommodation of layer mismatch. Relevance of electronic crystals for phase stability may be ranging from ceramics to intermetallics in an encompassing manner. Within 123, self-doping on CuO2 planes can lead to pairs of either electron or hole nature, accommodating the planes in a lattice that is in competition with chains concerning space requirements and fit. Charge on the chains will organize itself in a similar reciprocal way but generally in singles, safe for ETS. Nevertheless, it too will be employed in the process of healing of layer mismatch. This kind of structure stabilizing through self-doping appears to be of general applicability. Electronic crystal formation is therefore probably a widely utilized but yet unrecognized mechanism in structural chemistry. As such the case of superconductivity in ceramics appears to point out fundamental new criteria in a wider range of applications.

We assume that it may also be possible to extend the presumed 3-D phenomenology to other compound classes such as single layer Tl analogs. An extension of the ideas presented here is also possible for H-S and H-Se systems, which have shown Tc up to 190K [54] under pressure. They appear also to show the basic prerequisites of buckled planes allowing 3J doped bonding and may have the potential of 3-D coupling through 3J bonding also perpendicular to planes. In fact, they may be based on it. Generally, engineering plane expansions through substitutions may make materials susceptible to ETS. Examples would be RBa2Cu3Oy analogs with R=La or Pr which already display signs reminiscent of localization catastrophes that may be a precursor to ETS. Single layer deposition on templates with large inter-atomic metric such as the C- compound may be successful in achieving n-doping on planes. Although impractical, Ra substitution for Ba could also be employed in principle.

The more general open questions concerning ETS in 123 pertain to the nature of the 3-D interactions and the energetic of the cuprate self-doping reciprocity between plane and charge balance layer and its role on phase stability. In this respect one will have to learn to select rules amongst the competing trends for stabilizing ETS. As an example, the type of mechanisms need to be clarified that act in the competition between (3) and (4) leading to cell volume neutral intermediary states that are non-superconducting. This competition appears as a major negative influence on 2-D superconductivity and as a serious limit in the design of ETS. In this connection it is especially vexing to have at hand, in the P+C- phase at y=6.7, a stable material that shows the structural hallmarks of ETS but is non-superconducting. Moreover, it is recalcitrant to change its composition. One will also want to get a clearer understanding how the deleterious effect of d can be turned around and to advantage and how local modifications can be influenced based on O-O potentials. In other words, an encompassing crystal chemistry of all 3 types of cuprate superconductivity is now needed and within reach. But on a more practical side, a probably achievable challenge is now to stabilize Tc>300K with the strategies outlined.

**Summary**

Indications for above room temperature superconductivity are explained in terms of the creation of additional superconducting pairs on chains. As this is considered to result from plane expansions under n-doping, we propose material design that would foster these trends. Specifically for shot quenched materials a suitable mix of (n) appears required with d of doping-inactive (4) capable of supporting the self-doped oxidation of (2) in tandem with apical O, while (3) can be maintaining cell volume expansion. Laser pulsing of conventionally prepared YBa2Cu3Oy with y<6.5 such as z=0.32 with 68%(2) would be of interest for understanding limits, as would be the corresponding shot quenched materials. Materials with 32%(2) and c=0.080=2/5x5 are predicted to display Tc=352K* in shot quenched or laser pulsed materials. A P+C- compound at z=0.67 holds promise when prepared around z=0.5 under maintenance of its structural characteristics, as it is structurally related to laser pulsed materials.

**References**

1 H Oesterreicher, Solid State Communications 142 583 (2007)
2 H Oesterreicher, arXiv:0811.2792 (2008)

3 H Kotegawa et al, Phys. Review B 64 064515 (2001)
4 H Oesterreicher, Journal of superconductivity 16 (3), 507 (2003)
5 D Ko et al, Physica C: Superconductivity 277 95 (1997)
6 S Kaiser et al, Phys. Rev. B 89, 184516 (2014)
7 R Mankowsky et al, Nature 516 71 (2014)
8 H Oesterreicher, arXiv: 1511.09111 (2015)
9 H Oesterreicher, arXiv: 1512.02281 (2015)
10 AV Fetisov et al, Physica C: Superconductivity 515 54 (2015)
11 H Oesterreicher, Journal of superconductivity 16 (6), 971 (2003)
12 H Oesterreicher, Journal of superconductivity 17 (3), 439 (2004)
13 H Oesterreicher, Solid state communications 137 (4), 235 (2006)
14 H Oesterreicher, Journal of superconductivity 20 (3), 201 (2007)
15 H Oesterreicher, Physica C: Superconductivity 460 362 (2007)
16 H Oesterreicher, arXiv: 1503.00779 (2015)
17 T Hanaguri et al, Nature 430 1001 (2004)
18 KM Shen et al, Science 307 901 (2005)
19 Y Ohta et al, Phys. Rev. B 43 2968 (1991)
20 H Oesterreicher, Journal of superconductivity 14 (6), 701 (2001)
21 D Ko et al, Materials research bulletin 29 1025 (1994)
22 D Ko et al, Physica C: Superconductivity 231 252 (1994)
23 JR O'Brien et al, Journal of alloys and compounds 267 (1), 70 (1998)
24 H Oesterreicher et al, Journal of Alloys and Compounds, 269, 246 (1998)
25 H Oesterreicher et al, Journal of Alloys and Compounds 306 96 (2000)
26 HS Kim et al, Journal of Alloys and Compounds 339 65 (2002)
27 JR O'Brien et al, Physica C: Superconductivity 388, 379 (2003)
28 A Manthiram et al, Nature 329 701 (1987)
29 H Oesterreicher et al, Materials Research Bulletin 23 1327 (1988)
30 BE Higgins et al, Materials Research Bulletin 24 (6), 739 (1989)
31 MG Smith et al, Physical Review B 42 (7), 4202 (1990)
32 MG Smith et al, Journal of Applied Physics 69 (8), 4894 (1991)
33 MG Smith et al, Physical Review B 46 (5), 3041 (1992)
34 MG Smith et al, Journal of Solid State Chemistry 99 140 (1992)
35 LT Romano et al, Physical Review B 45 (14), 8042 (1992)
36 H Oesterreicher Applied physics 15 (4), 341 (1978)
37 D Ko et al, Physica C: Superconductivity 231 (3), 252, (1994)
38 D Ko et al, Materials research bulletin 29 (10), 1025 (1994)
39 H Oesterreicher, Materials research bulletin 30 (8), 987 (1995)
40 M Scavini et al, Europhys. Lett. 76 443 (2006)
41 G Ceder et al, Phys. Rev. B., 44 2377 (1991)
42 HF Poulsen et al, Phys. Rev. Lett., 66 465 (1991)
43 R McCormack et al, Phys. Review B., 45 12976 (1992)
44 DJ Liu et al, Phys. Rev. B., 52 9784 (1995)
45 F Yakhou et al, Physica C., 333 146 (2000)
46 M Zimmermann et al, Phys. Rev. B., 68 104515 (2003)
47 R Liang et al, Phys. Review B., 73 180505 (2006)
48 H Oesterreicher, Journal of alloys and compounds 267 (1), 66 (1998)
49 H Oesterreicher, Journal of alloys and compounds 319 (1), 131 (2001)
50 NP Armitage et al, arXiv:0906.2931v2 (2010)
51 B Keimer et al, arXiv: 1409.4673 (2014)
52 H Oesterreicher, Journal of Superconductivity 18 509 (2005)
53 S Gerber et al, Science 350 949 (2015)

54 AP Drozdov et al, arXiv: 1412.0460 (2014)

Table 1. Structural and electronic types in YBa2Cu3Oy and comparison compounds indicating 3 categories P-, P+, and ETS/C-. P+ stands for plane expansion and self-doping of n-type, relative to the semiconductor denoted P0. P+ can balance C- effects, which stands for chain contraction through double p-doping. P+ is characteristic for quenched elevated temperature states while the opposite self-doping of P- type can develop when low temperature shrinks the planes.
In the definition c=n/Pab, n is the number of pairs per plaid, and Pab is the period of bond order in multiples of lattice parameter a0 and b0. For the pairs only source-region between the bond orders 5a0x5b0 and 3a0x3b0, Tc=ic where i is a plane isolation factor that can be calculated from apical O distance d. Star indicates a situation where the total number of pairs has actually increased through additional pairs on (2). The resulting ETS should be favored by large plane metric as achieved on laser pulsing and shot quenching.
One can write charge balance for conventional P- O24 YBa2Cu3O6.5 (5)2.125+, (4)2.75+, (2)1+, while for laser pulsed P+C- O24 YBa2Cu3O6.5 it can be assumed as (5)1.875+, (2)1.25+, (4)3+. For shot quenched P+ O32 YBa2Cu3O6.44 it is (5)1.78+, 89%(3)2.44+, 11%(2)1+, which can transform to conventional P-. The transitory product 22%(2), 67%(3), 11%(4) may be the origin of Tc=200K* level with (2) as an n-dopant and a calculated Tc=242K*.
Compound P+C- T34 YBa2Cu3O6.67 is structurally related to laser pulsed YBa2Cu3O6.5 of ETS type although it is non-superconducting, tetragonal and assumed to be composed of 67% (3) and 33% (4).
Superscripts o and c indicate observed and calculated. Sign ‘ indicates singular or uncertain value.

| c=n/Pab | Type, examples | Tc level | Treatment |
|---|---|---|---|
| 0.25=4/4x4 | P+C- T34 YBa2Cu3O6.67 | - | quench |
| 0.222=2/3x3 | P- O42 YBa2Cu3O6.89 | 90 | conventional |
| 0.222*=2/3x3 | P+ O32 YBa2Cu3O6.44’ | 200* | quench |
| 0.222=2/3x3 | P+ O32 YBa2Cu3O6.44 | 100 | quench |
| 0.167=2/3x4 | P- O42 YBa2Cu3O6.67 | 67 | conventional |
| 0.125=2/4x4 | P- O24 YBa2Cu3O6.5 | 50 | conventional |
| 0.125*=2/4x4 | P+C- O24 YBa2Cu3O6.5 | 550* | laser |
| 0.080=2/5x5 | P- O24 YBa2Cu3O6.32 | 30 | conventional |
| - | P0C0 T23 YBa2Cu3O6.25 | - | conventional |

Table2. Predicted Tc=%Dxconst, where D stands for self-dopant, in our particular case (2), and const is calibrated as 11 for assumed 3-dimensional superconductivity (ETS). Shown are slow cooled O24 emphasizing predictions for laser pulsing. Also shown are examples of a mix with (3)-phenomenology based on $O^-$. The latter, prepared at elevated temperatures, have potential to show ETS in the bulk. Stabilizing factors for the requisite thick family properties (C-P+) include supporting mix of (n) and charge lattice commensurability of self-doped charge. Examples of (3)-phenomenology are expected with %(2) of 67, 50, and 32 with magic number counts on planes of c= 0.167, 0.125, and 0.080 respectively, due to (2) self-doping. They correspond to z of 0.16, 0.25, and 0.34. Tc accordingly are predicted as 737K*, 552K*, and 352K*.

| Examples | %(2) | %(3) | Tc(K) |
|---|---|---|---|
| YBa2Cu3O6.89 O24 | 11 | 0 | 121* |

| | | | | |
|---|---|---|---|---|
| YBa2Cu3O6.72 | O24 | 28 | 0 | 308* |
| YBa2Cu3O6.68 | O24 | 32 | 0 | 352* |
| YBa2Cu3O6.50 | O24 | 50 | 0 | 550* |
| YBa2Cu3O6.50 | O324 | 25 | 50 | 275* |
| YBa2Cu3O6.44 | O32 | 11 | 89 | 121* |
| YBa2Cu3O6.44 | O324 | 22 | 68 | 242* |
| YBa2Cu3O6.44 | O234 | 50 | 39 | 550* |
| YBa2Cu3O6.44 | O24 | 56 | 44 | 616* |
| YBa2Cu3O6.33 | O24 | 67 | 0 | 734* |
| YBa2Cu3O6.32 | O32 | 32 | 68 | 352* |
| YBa2Cu3O6.25 | T23 | 50 | 50 | 550* |
| YBa2Cu3O6.06 | T23 | 89 | 11 | 981* |
| YBa2Cu3O6.00 | T2 | 100 | 0 | 1100* |